# SMART/SDDI Filament Disappearance Catalogue

**Daikichi SEKI**[1, 2, 3, 4], **Kenichi Otsuji**[1], **Takako T. Ishii**[1], **Kumi Hirose**[1], **Tomoya Iju**[5], **Satoru UeNo**[1], **Denis P. Cabezas**[1], **Ayumi Asai**[1], **Hiroaki Isobe**[6], **Kiyoshi Ichimoto**[1], and **Kazunari Shibata**[1]

1. Astronomical Observatory, Kyoto University, Japan
2. GSAIS*, Kyoto University, Japan
3. DAMTP**, University of Cambridge, UK
4. CSER***, University of Cambridge, UK
5. Solar Science Observatory, National Astronomical Observatory of Japan, Japan
6. Faculty of Fine Arts, Kyoto City University of Arts, Japan;

*Graduate School of Advanced Integrated Studies in Human Survivability
** Department Applied Mathematics and Theoretical Physics
*** Centre for the Study of Existential Risk

E-mail (seki@kwasan.kyoto-u.ac.jp)

**Abstract**

This paper describes a new "SMART/SDDI Filament Disappearance Catalogue," in which we listed almost all the filament disappearance events that the Solar Dynamics Doppler Imager (SDDI) has observed since its installation on the Solar Magnetic Activity Research Telescope (SMART) in May 2016. Our aim is to build a database that can help predict the occurrence and severity of coronal mass ejections (CMEs). The catalogue contains miscellaneous information associated with filament disappearance such as flare, CME, active region, three-dimensional trajectory of erupting filaments, detection in Interplanetary Scintillation (IPS), occurrence of interplanetary CME (ICME) and Dst index. We also provide statistical information on the catalogue data. The catalogue is available from the following website: https://www.kwasan.kyoto-u.ac.jp/observation/event/sddi-catalogue/.

1. **INTRODUCTION**

Dark filaments observed on the solar disk (or prominences on the solar limb) are comprised of dense cool plasma floating in the solar corona supported by a magnetic field. Their electron densities and temperatures are estimated as $10^9$–$10^{10}$ cm$^{-3}$ and around $10^4$ K, respectively (Parenti 2014). At the end of their lives, filaments sometimes disappear (known as "disparition brusque") or dynamically erupt, which is associated with a coronal mass ejection (CME). In this paper, we define "filament disappearance" as an event in which a filament disappears completely in an H-alpha line observation. An eruption is believed to be triggered by magnetic reconnection, ideal magnetohydrodynamic (MHD) instability or loss of equilibrium (Priest and Forbes 1990). The reconnection-induced triggering mechanism includes magnetic reconnection between the pre-existing flux rope containing the filament and the nearby emerging magnetic flux (Kusano, et al. 2012; Feynman and Martin 1995; Chen and Shibata 2000), the overlying fields above the flux rope (Antiochos, DeVore and Klimchuk 1999) or the underlying sheared fields below the filament (Moore, et al. 2001). As for the instability-induced triggering mechanism, several MHD instabilities including Torus instability (Kliem and Török 2006) and Kink instability (Sakurai 1976; Török and Kliem 2005) have been proposed.

Space weather is a plasma disturbance in the interplanetary medium and near-Earth space caused by solar magnetic activity. Recently, space weather has attracted much attention due to its potential societal and economic impacts (National Research Council 2008; Royal Academy of Engineering 2013). Since filaments often erupt and become a part of CMEs, they can also be a major cause of space weather. McAllister, et al. (1996) reported a polar crown filament eruption on 1994 April 4 and its associated geomagnetic storm (Dst ~ -200 nT) three days later. Cliver, et al. (2009) also reported a quiescent filament eruption event on 1991 November 9, which caused a significant geomagnetic storm (Dst ~ -354 nT) around three days after the eruption.

To mitigate the impacts of geomagnetic storms, a number of studies have sought to predict CME arrival times and the occurrence of severe geomagnetic storms (Gopalswamy et al. 2000; 2001; Liu et al. 2018) using coronagraphs, like Large Angle and Spectrometric Coronagraph (LASCO; Brueckner et al. 1995) onboard the Solar and Heliospheric Observatory (SOHO; Domingo, Fleck and Poland 1995). However, coronagraphs only provide the information on the plane-of-sky velocity of CMEs (i.e., they lack information on the line-of-sight velocity [LOSV]).

In an effort to compensate for the LOSV information of CMEs, in this paper, we present a catalogue of filament disappearances observed by the Solar Dynamics Doppler Imager (SDDI; Ichimoto et al. 2017) installed on the Solar Magnetic Activity Research Telescope (SMART; UeNo et al. 2004) at Hida Observatory, which is called the "SMART/SDDI Filament Disappearance Catalogue." SDDI has been used to observe the solar full-disk image since 2016 May 1 in 73 wavelengths around the H-alpha absorption line at 656.3 nm. The time cadence is 12 seconds, and the spatial sampling is 1.23 arcsec pix$_{-1}$. SDDI's key benefits are its continuous full-disk observation of the Sun and its wide coverage of observing wavelengths. It captures the full-disk Sun from H-alpha centre − 9.0 Å (blue shift) to H-alpha centre + 9.0 Å (red shift) every 0.25Å. The high resolution in wavelengths and the wide coverage around H-alpha line enables us to determine the LOSV of filaments in unprecedented detail due to its dynamic range of ±400 km s$_{-1}$. For more details on SDDI, see Ichimoto et al. 2017.

## 2. Overview of the Catalogue

The catalogue webpage can be accessed at https://www.kwasan.kyoto-u.ac.jp/observation/event/sddi-catalogue/. Figure 1 provides an overview of the catalogue arranged as a table. The table lists all the filament-disappearance events observed by SDDI from 2016 May 1 to 2019 June 18. To identify events, we first searched for the word "filament" in the daily observation logs created by human operators. Then, if eruptive events were reported on a particular day, we investigated whether there was a filament disappearance in the H-alpha centre on that day by investigating the daily quick-look movie. The catalogue contains 36 columns related to flare association, CME association, filament disappearance dynamics, interplanetary conditions and geomagnetic activity. Each column is described in the following section.

Figure 1 Table layout.

## 3. Column descriptions

The first two columns, "id & obs_date" and "event_reports," identify filament disappearance events with a number and observing date. Clicking the id number takes to a full-disk image of the Sun taken in the H-alpha line centre (see Figure 2). Clicking the observation date takes to the webpage for SDDI daily observations, which include a java movie and fits files of the day's observations.

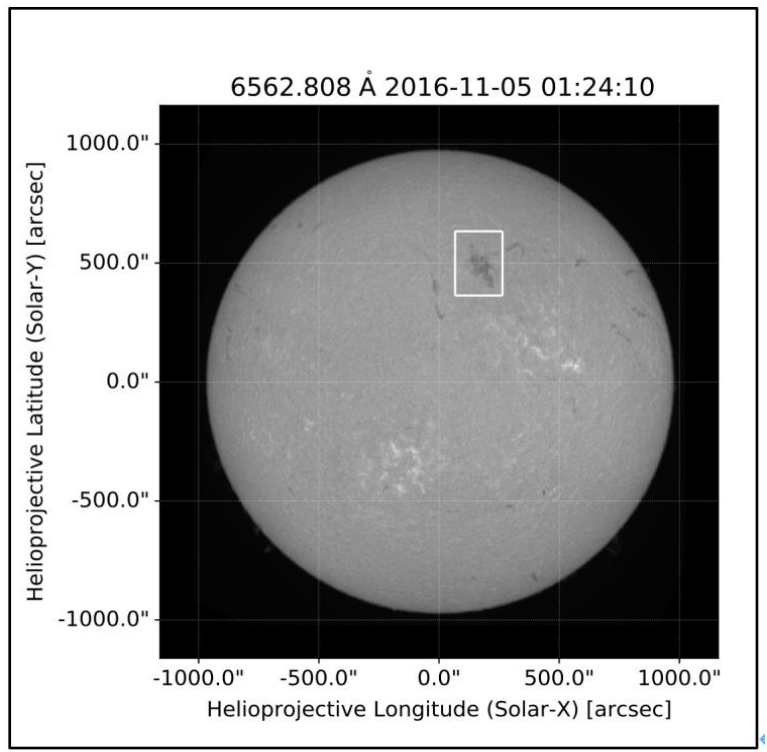

Figure 2. A full-disk image of the Sun in the H-alpha line centre captured when a filament started to disappear (same as "FD_start_time"). The white rectangle encloses the target filament.

The next five columns, "flare_class," "flare_peak_time," "noaa_ar," "ar_location" and "ar_size," provide information on the associated flare. "Flare_class" and "flare_peak_time" correspond to the soft X-ray flare class peak time from the Geostationary Operational Environmental Satellite (GOES) observations. Clicking these values reveals the soft X-ray light curve. The "noaa_ar," "ar_location" and "ar_size" are the NOAA Active Region (AR) number, the AR location in heliographic coordinates, and the rough AR size, respectively. Note that the AR size is measured as the size of the rectangle containing the entire AR, corrected for projection effects due to sphericity. Clicking these values takes to the corresponding webpage in the Solar Monitor (https://www.solarmonitor.org).

The next five columns, "CME," "central_PA," "angular_width," "linear_speed" and "credibility," provide information on the associated CME. "CME," "central_PA," "angular_width" and "linear_speed" correspond to the date and time of the CME's first appearance in the LASCO C2 field of view, the central position angle, the angular width and the CME's linearly estimated speed, listed in the SOHO/LASCO CME Catalog (hereafter the "CME Catalog"; Yashiro et al. 2004; Gopalswamy et al. 2009). As the CMEs in the CME Catalog were manually tracked, some CMEs (particularly very poor events) were not listed in the CME Catalog. In such cases, we tracked the LASCO images by ourselves. In the "CME" column, we added a comment "-" if there was no CME candidate, or "nan" was input if there was no investigation into associated CMEs, mainly due to a lack of LASCO data. Clicking these four values takes to the corresponding webpage in the CME Catalog. "Credibility" denotes the subjective credibility of the CME association. Higher values indicate higher credibility; events with a credibility of 1 may be controversial. Clicking the credibility shows the actual movies that were used to investigate the CME association (see Figure 3): the solar full-disk movie in H-alpha (for most of the events) or in 304 Å captured by the Atmospheric Imaging Assembly (Lemen, et al. 2012) onboard the Solar Dynamics Observatory (Pesnell, Thompson and Chamberlin 2012) and the running difference movie of SOHO/LASCO C2 are shown.

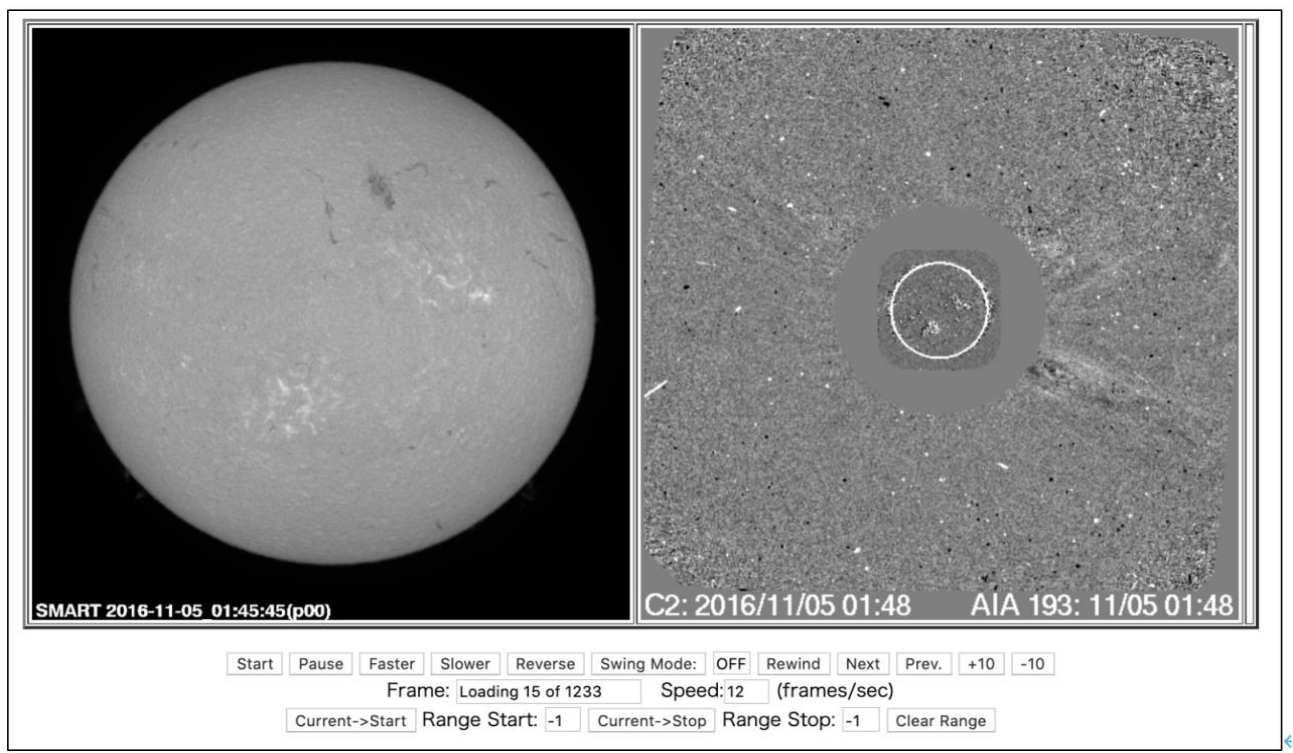

Figure 3. A snapshot of the two movies that were utilized to investigate the association between a filament disappearance and a CME.

The next sixteen columns, "FD_start_time," "FD_end_time," "x," "y," "longitude," "latitude," "Vx," "Vy," "Vz," "Vr_max_time," "Vr_max," "Vr_fin," "tracked_data," "phi," "theta" and "inclination_angle," pertain to filament dynamics. "FD_start_time" and "FD_end_time" correspond to the start and end times of a filament disappearance, which are defined as the time when the filament appeared in the H-alpha line centre – 0.5 Å and the time when the filament totally disappeared from all observed wavelengths, respectively. Clicking these two values takes to the corresponding movie, a snapshot of which is shown in Fig. 4.

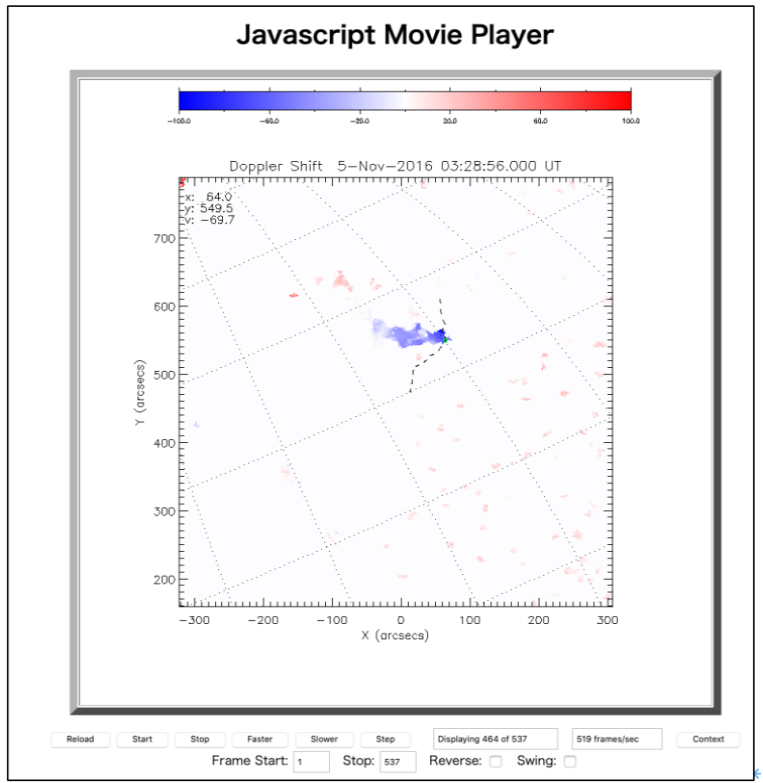

Figure 4 Snapshot of the LOSV movie and the trajectory of a filament. The black dashed line denotes the trajectory of the filament.

The black dashed line indicates the trajectory of the filament, along which we measured the X and Y positions of the filament. The small green rectangle corresponds to the measured position at that time. The measured position was determined as the apex point observed in "weighted averaged contrast (WAC)" image, which is the average contrast through SDDI wavelengths (H-alpha centre – 9.0Å to the H-alpha centre + 9.0Å) with specific weights. The value of WAC, $I_{wac}$, was calculated by

$$I_{wac} = \frac{1}{N} \sum_{\widehat{C_\lambda} > 3.0} \widehat{C_\lambda}$$

$$\widehat{C_\lambda} = \frac{C_\lambda}{S_\lambda}$$

and

$$C_\lambda = \frac{I_\lambda - I_{0\lambda}}{I_{0\lambda}}$$

where $N$ is the number of wavelengths in which $\widehat{C_\lambda}$ is larger than 3.0, $I_\lambda$ and $I_{0\lambda}$ are intensities of the target filament and background evaluated from the surrounding regions at the wavelength $\lambda$, and $S_\lambda$ is the weight value shown in Figure 5, which empirically corresponds to the standard deviation of contrast, $C_\lambda$. The WAC image can collect the darker features from all wavelengths, which allows tracking the target filament, regardless of the Doppler shift (for more details, in Otsuji et al., in prep.). Figure 5 is a snapshot of an actual WAC movie constructed from SDDI observations on 2016 November 5. Note that links to WAC movies are not included in the catalogue. The WAC movie can be accessed at *https://www.kwasan.kyoto-u.ac.jp/~otsuji/trajectory_data/{"FD_start_time" in YYYYMMDDHHmm00}/proc0/movie.html* (see Figure 5).

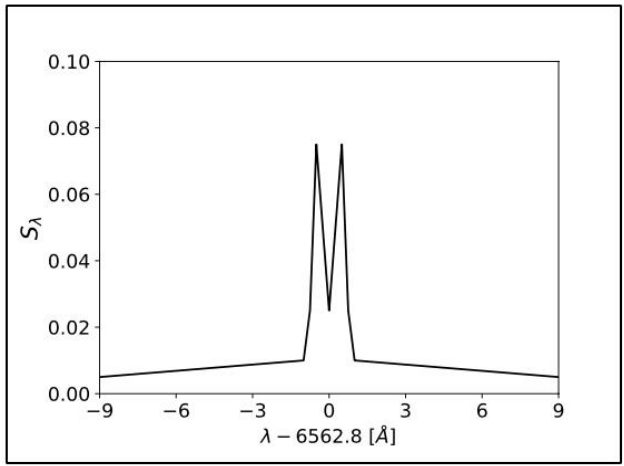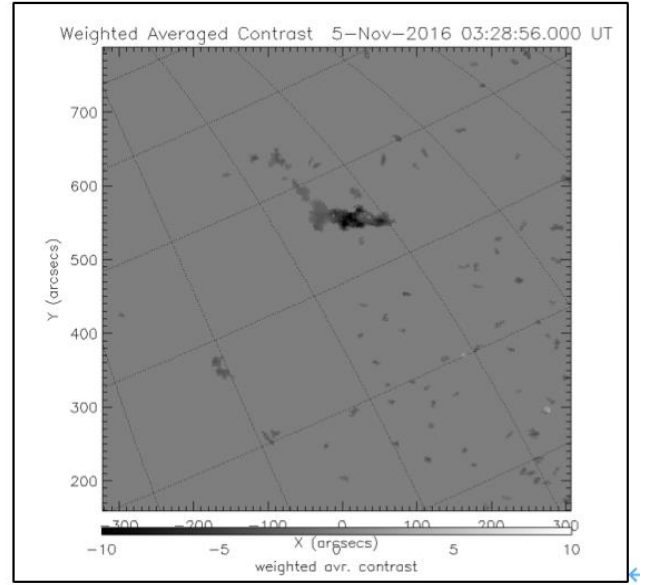

Figure 5 (left) Weight value $S_\lambda$ along wavelengths. (right) A snapshot of WAC movie on 2016 November 5.
(https://www.kwasan.kyoto-u.ac.jp/~otsuji/trajectory_data/20161105012400/proc0/movie.html)

"X," "y," "longitude" and "latitude" indicate the position of the approximate centre of the filament at "FD_start_time". Clicking these four values shows the solar full-disk image and the target filament (same as clicking "id"). Note that the projection effects, due to the extended height of a filament, were not taken into account, leading to a few degrees of uncertainty in "longitude," "latitude" and "inclination angle". "Vx," "Vy," "Vz," "Vr_max_time," "Vr_max" and "Vr_fin" correspond to the representative velocity components of the filament and measured time. Firstly, we manually measured the filament's X- and Y-positions and LOSV. Then, assuming the three components of its position were 0 Mm at the beginning of measurement (same as "FD_start_time") and determining the LOS location by integrating its LOS velocity with the time, we constructed its position in the heliocentric Cartesian coordinates (see Figure A in the webpage). We also transformed and obtained the position in the local Cartesian coordinates as well, in which three orthogonal base vectors were directed along the solar normal, latitudinal direction and longitudinal direction. The maximum radial velocity component was obtained as the maximum derivative of the interpolated spline from the radial positions. It was listed as "Vr_max" and its measured time was "Vr_max_time." Three velocity components in the heliocentric Cartesian coordinates at that time, "Vx," "Vy" and "Vz," were derived in the same manner. "Vr_fin" corresponds to the radial velocity at the "FD_end_time." Clicking these values shows the time-distance plots in two different coordinates (see Figure 6), the heliocentric Cartesian coordinates and the local Cartesian coordinates. "Tracked_data" contains the actual tracked locations in csv format. For more details on how to measure these velocity components, see Otsuji et al. (in prep). "Phi" and "theta" indicate the velocity angles in spherical coordinates. "inclination_angle" shows the angle between the local solar normal and the velocity (see Figure A and B in the webpage). Clicking these angles demonstrates the filament's three-dimensional trajectory (see Figure 7)

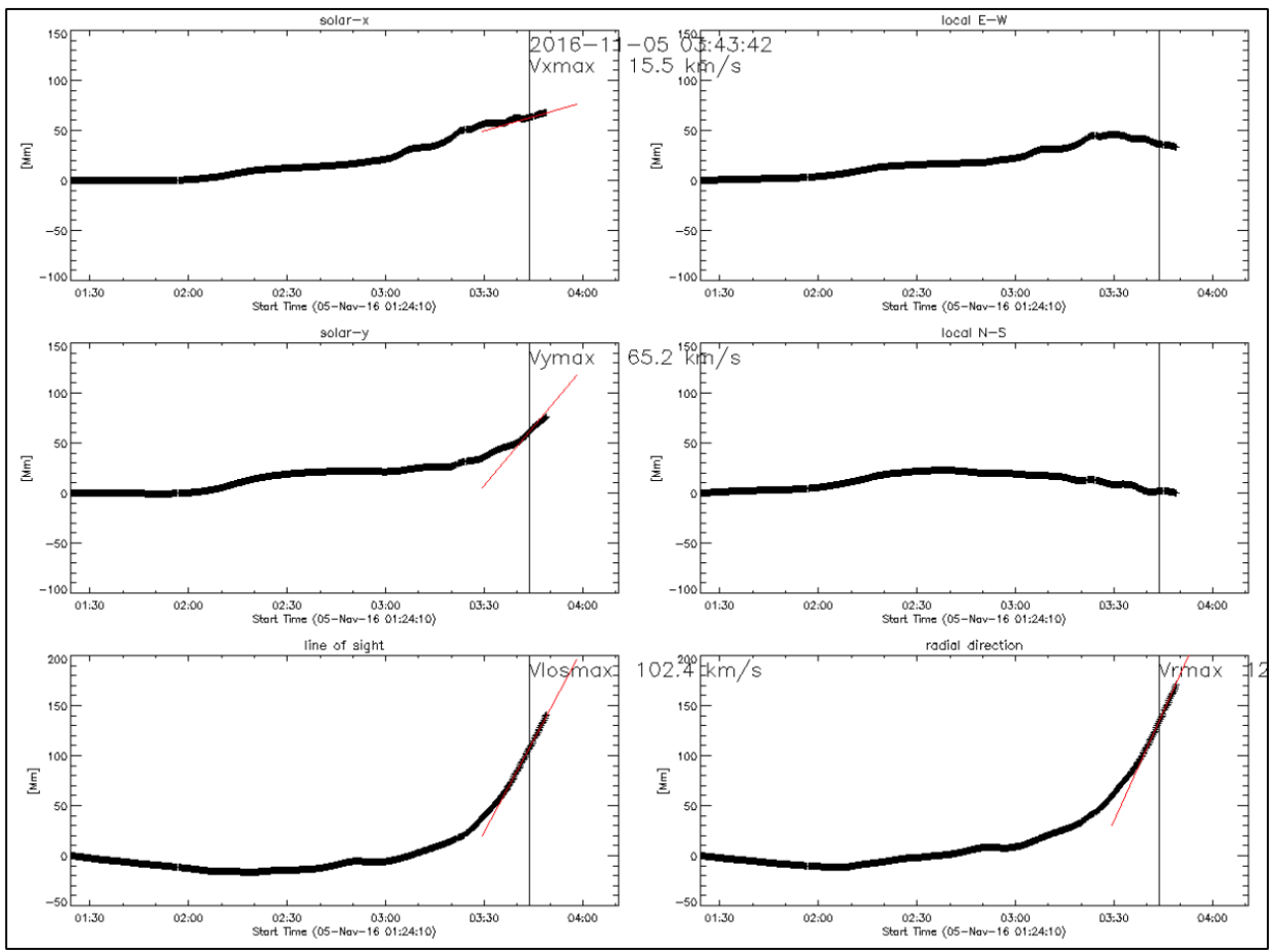

Figure 6 (Left) Time-distance plots in heliocentric Cartesian coordinates, in which x- and y-axis are directed to the solar west and north, respectively. Z-axis points toward the Earth along the line-of-sight direction. (Right) Same plots in local Cartesian coordinates. Local E-W and local N-S indicate the longitudinal and latitudinal directions on the solar surface, respectively. The black and red lines indicate the time when the radial velocity was maximum and the tangent of the height-time plot, respectively.

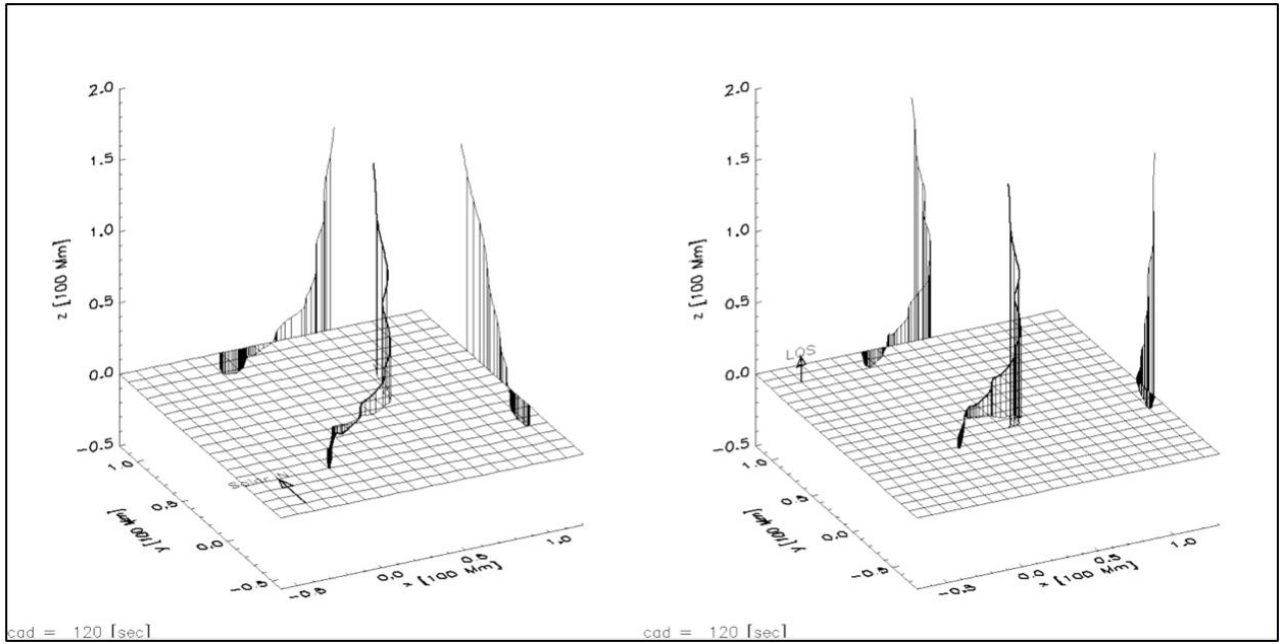

Figure 7 Three-dimensional trajectory of the filament. The left and right panels exhibit the trajectories in the heliocentric Cartesian coordinates (z-direction is LOS) and the local Cartesian coordinates (z-direction is the solar normal). The arrows in the left and right panels point to the solar north and to the Earth, respectively. The black vertical lines between three-dimensional trajectories and the horizontal planes (z = 0) are drawn every two minutes.

The three columns, "length," "length_csv" and "Type," are related to the filament's length and type before its disappearance. The length was measured manually as shown in Figure 8. Clicking the "length" displays the same image for each event. The time series of the measured positions is stored in the csv file, "length_csv". Each filament or prominence was classified as one of the following three types, based on the connectivity to the active regions; active region filament (AF), intermediate filament (IF) and quiescent filament (QF). If the filament was entirely in the active region, it was classified as AF. If there was no nearby active region, it was QF. If the filament was neither AF nor QF, it was categorized as IF.

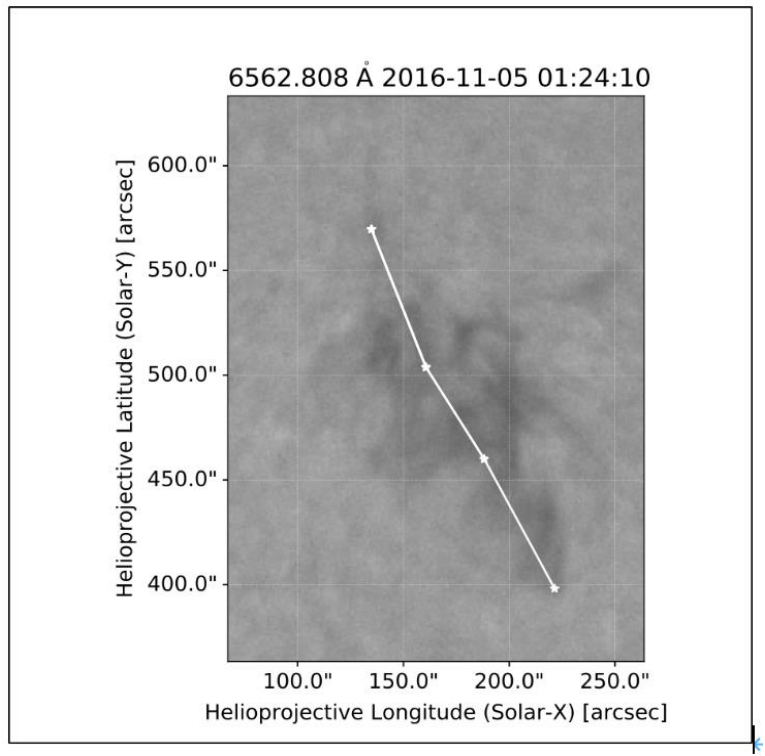

Figure 8. Close-up image of the target filament. This white line indicates trace of the filament when the length was measured.

The next column, "IPS," corresponds to the g-values of the associated enhancement of the interplanetary scintillation (IPS) and its observed date. IPS is the disturbance of the radio wave from a distant source caused by the interplanetary plasma, such as from the solar wind or CMEs. The g-value is defined as the ratio of the IPS disturbance at the observed time to its annual average. For more details about IPS and g-value, see Gapper et al. (1982) and Iju et al. (2013). For all CME-associated events, we investigated whether the g-value was greater than 1.5 around the equivalent position angle of the CME. The statistical mean and standard deviation of all the g-values, observed from 1997 to 2009, were 1.07 and 0.47, respectively. Thus, we regarded a g-value greater than 1.5 as indicating the presence of interplanetary plasma cloud. Placing a cursor on the date shows the g-values greater than 1.5. Clicking the date displays the daily chart of the IPS observation, as shown in Figure 9.

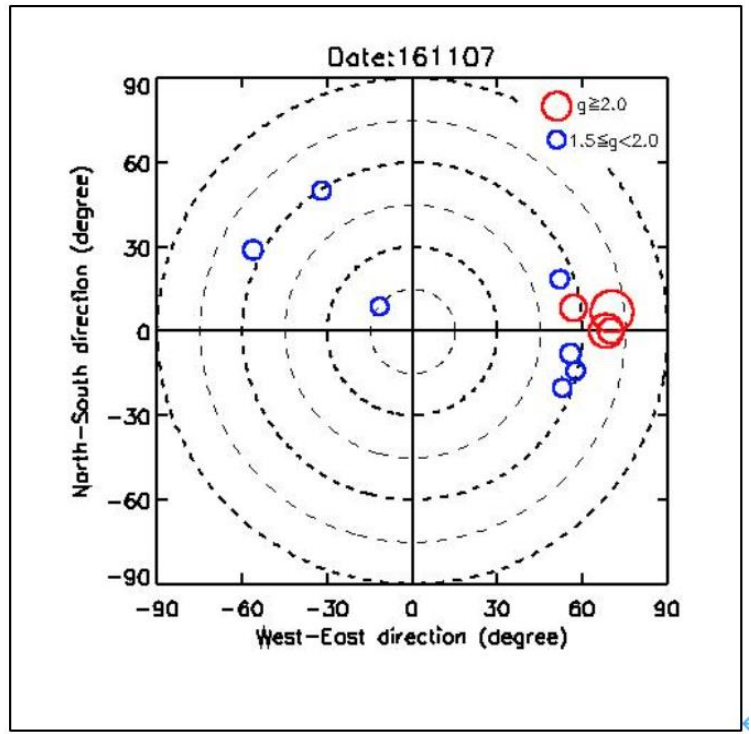

Figure 9. Daily IPS observation. The blue and red circles point to the positions where g-values are 1.5–2.0 and ≥ 2.0, respectively. The center of the map corresponds to the Sun.

The column "ICME" lists the arrival time of the associated interplanetary CME registered in the Richardson & Cane ICME Catalog (http://www.srl.caltech.edu/ACE/ASC/DATA/level3/icmetable2.html). Interplanetary CMEs (ICMEs) were detected as an abnormally low temperature plasma, the characteristics of which are common for most of CMEs. For more details, see Cane and Richardson (2003).

The two columns, "Dst_peak_time" and "Dst_peak_value," indicate the time and peak of the Dst index1–7 days after the filament disappeared. Note that this peak value is not necessarily associated with the filament disappearance. Users should be careful when considering the causes of geomagnetic disturbance. Clicking these values shows the monthly plot of the Dst index at the World Data Center for Geomagnetism, Kyoto. The last column notes a comment. Placing a cursor over the word, "note," shows it as a popover.

## 4. Statistical properties

In this section, we provide information on the statistical properties of the filament disappearance events in our table. Figure 10 demonstrates the distribution of the maximum radial velocities. The averages for all the filament disappearances, CME-associated and CME-unassociated disappearances were 124, 177 and 55.6 km s$_{-1}$, respectively. 59% of the filament disappearances with their maximum radial velocities <140 km s$_{-1}$ were not associated with CMEs and 90% of the filament disappearances with their maximum radial velocities >140 km s$_{-1}$ were associated with CMEs. The faster the filaments disappeared in the radial direction, the more likely they are associated with CMEs. Table 1 summarizes statistical properties of Figure 10. The value in a parenthesis corresponds to the number of the clear events whose values of "credibility" are 2 or 3.

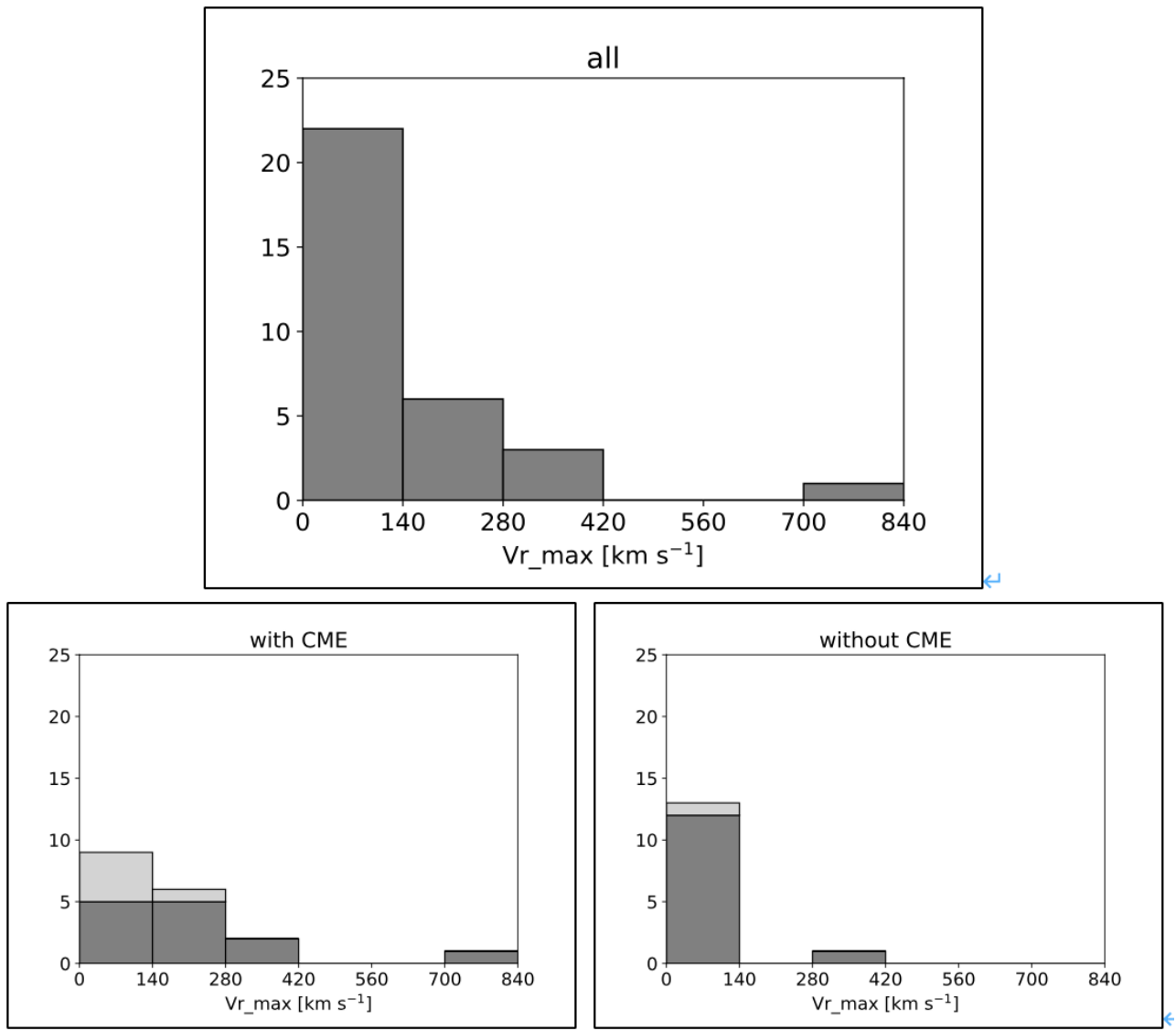

Figure 10  Distribution of maximum radial velocities of all the filament disappearances (top), of those associated with CMEs (bottom left), and those without associated CMEs. Light gray bins correspond to the events whose values of "credibility" are 1 (i.e. unclear events).

Table 1 Statistical properties of the maximum radial velocities.

|  | all events | with CME | without CME |
|---|---|---|---|
| < 140 km s$^{-1}$ | 22 | 9 (5) | 13 (12) |
| > 140 km s$^{-1}$ | 10 | 9 (8) | 1 (1) |
| total | 32 | 18 (13) | 14 (13) |

Figure 11 shows the distribution of the inclinations between the maximum velocity of erupting filaments and LOS direction. The average theta angle was 64.4 deg. For 26% of the events, the theta angles were larger than 90 deg, which corresponds to an opposite direction to the Earth.

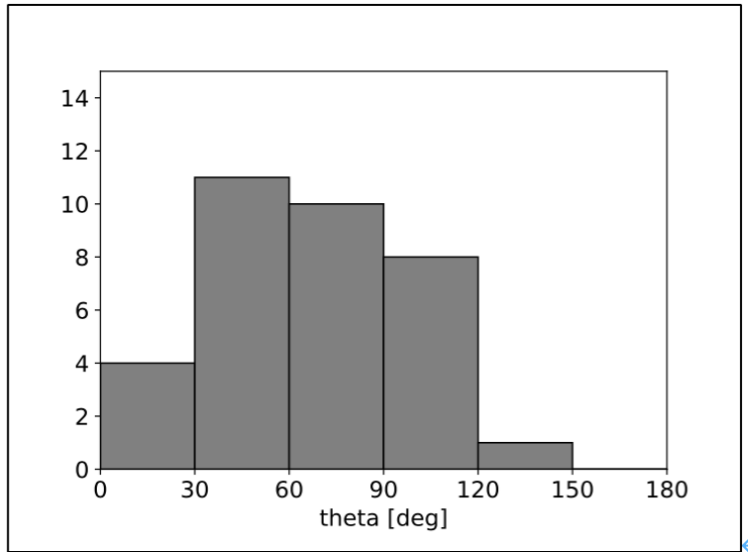

Figure 11 Distribution of theta angles of filament disappearances with respect to the LOS direction.

Figure 12 demonstrates the distributions of the inclinations between the direction of an eruption and the solar normal. The histogram is displayed in terms of the solid angle derived by "inclination_angle" in the catalogue. The average solid angle is 2.10 str which corresponds to 48.2 deg in terms of "inclination_angle". For the events whose solid angles are less than $\frac{2\pi}{3}$, 65% of filament disappearances are associated with CMEs. For those with their solid angles more than $\frac{2\pi}{3}$, 58% of filament disappearances are not associated with any CMEs. Only taking the clear events into account, we can see that 75% of filament disappearances are not associated. This result is consistent with Gopalswamy et al. (2003), in which transverse events, whose solid angles are expected to be large, are less associated with CMEs than radial events, whose solid angles should be small. Table 2 summarizes statistical properties of Figure 12. The value in a parenthesis corresponds to the number of clear events.

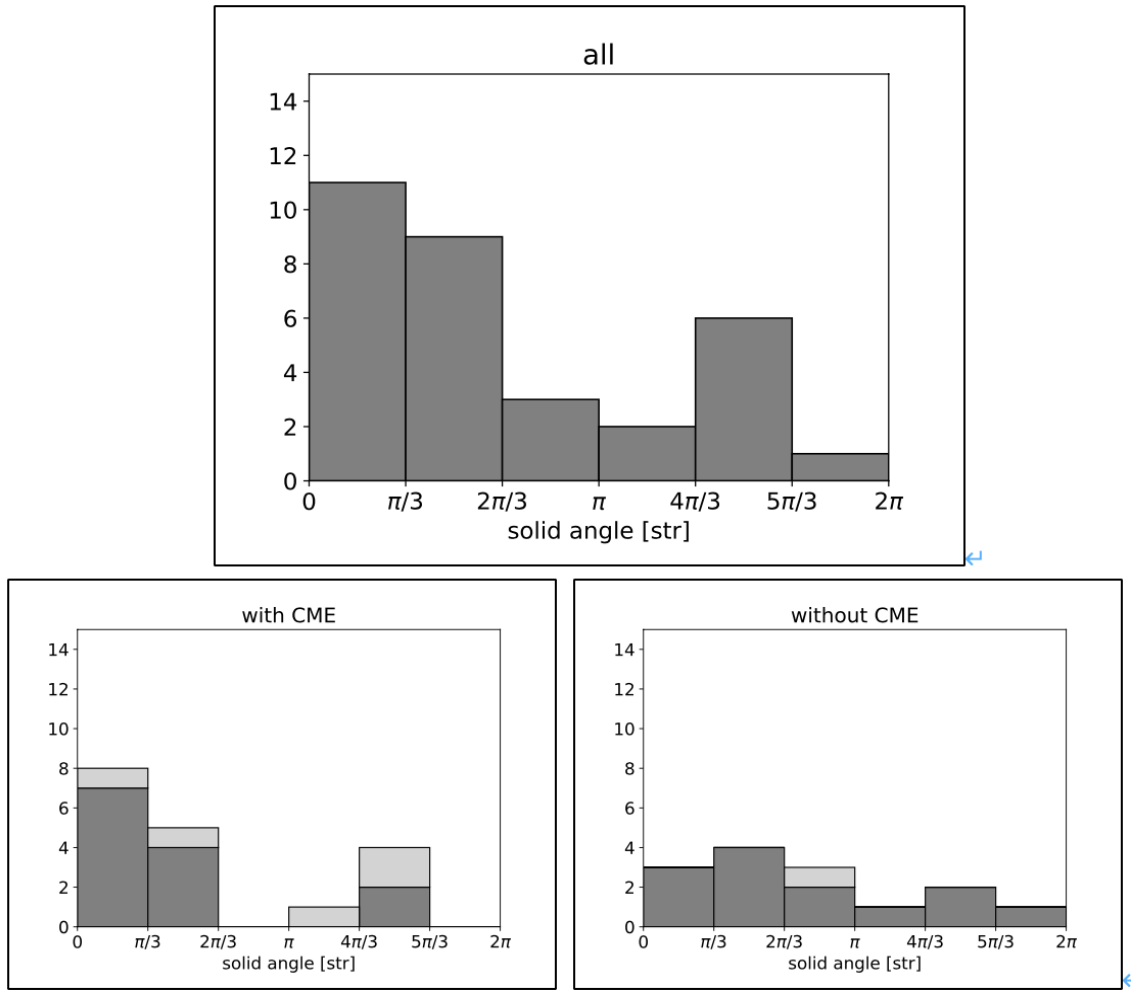

Figure 12 (top) Distribution of inclinations of filament disappearances in terms of solid angle derived by "inclination_angle". (bottom left) only for the events associated with CMEs. (bottom right) only for events without CMEs.

Table 2 Statistical properties of the solid angles.

|  | All events | with CME | without CME |
|---|---|---|---|
| $< \frac{2\pi}{3}$ str | 20 | 13 (11) | 7 (7) |
| $> \frac{2\pi}{3}$ str | 12 | 5 (2) | 7 (6) |
| total | 32 | 18 (13) | 14 (13) |

Figure 13 shows the distributions of the apparent lengths of filaments at the beginning of their disappearances ("FD_start_time" in the catalogue). The average length was 139 Mm. 63% of the filaments whose lengths are less than 150 Mm are not associated with CMEs during their disappearances. On the other hand, 67% of those with their lengths of more than 150 Mm disappeared with CMEs. We can see that the longer the filaments, the more likely they are to be associated with CMEs. Table 3 summarizes the statistical properties of Figure 13. The value in a parenthesis corresponds to the number of clear events. It must be noted that some narrow CMEs from close to the disk center may not be observed by LSACO, so the lack of association may be due to this visibility effect in at least some cases.

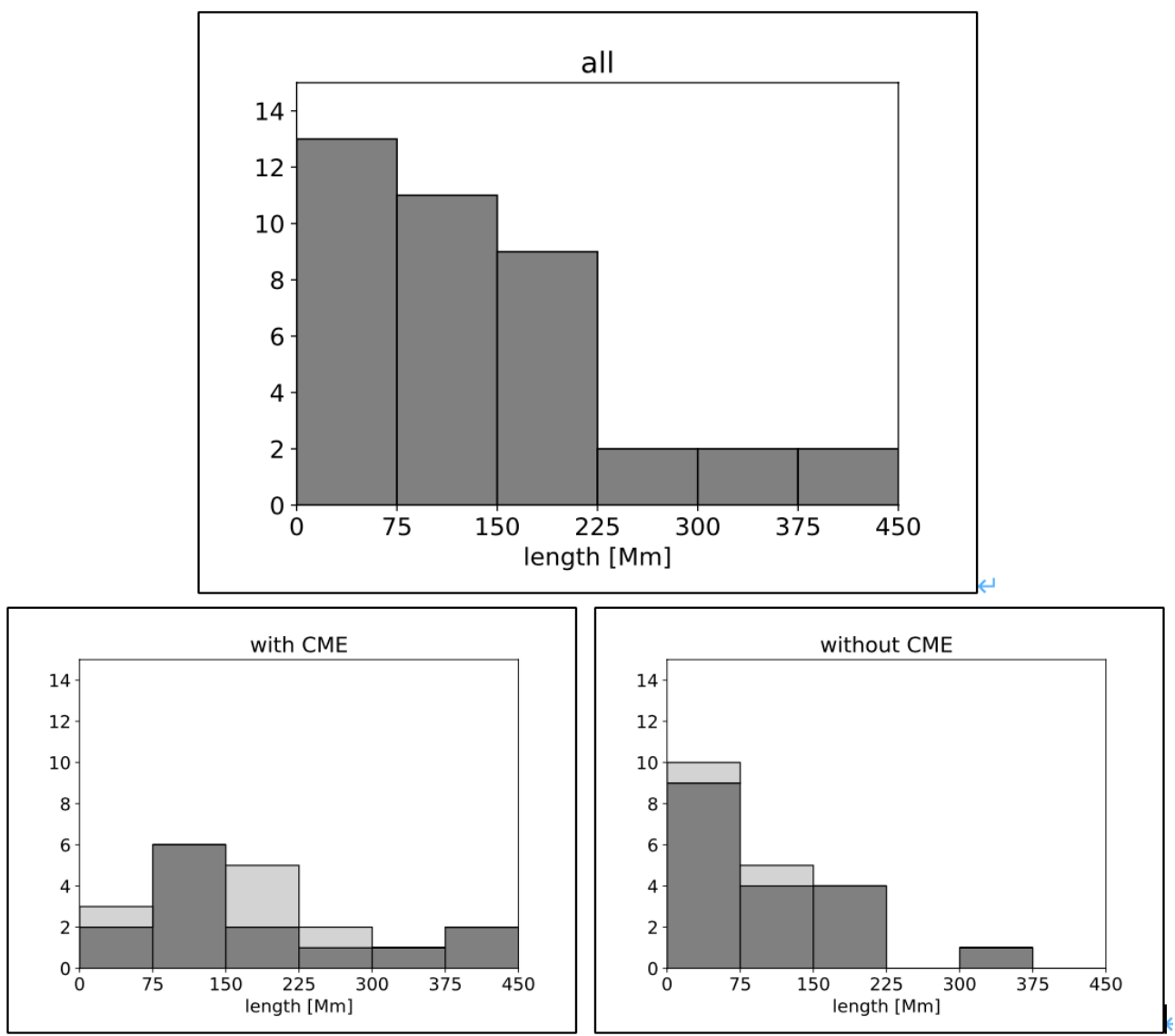

Figure 13 (top) Distribution of apparent lengths of all the filaments. (bottom left) Only for those associated with CMEs. (bottom right) Only for those without CMEs.

Table 3 Statistical properties of the lengths.

|           | all events | with CME | without CME |
|-----------|------------|----------|-------------|
| < 150 Mm  | 24         | 9 (8)    | 15 (13)     |
| > 150 Mm  | 15         | 10 (6)   | 5 (5)       |
| total     | 39         | 19 (14)  | 20 (18)     |

## 5. Summary

This paper presented the "SMART/SDDI Filament Disappearance Catalogue" that aims to investigate essential parameters for predicting CMEs' arrival times and the potential geomagnetic impacts. The catalogue listed 43 filament-disappearance events observed by SDDI since the beginning of its operation (2016 May 1) with miscellaneous information, including the associated flare; active region; CME; the position, dynamics and properties of the filaments; IPS; and geomagnetic disturbance. Statistical properties were provided as well and we recognize that the faster filaments disappeared, the less inclined to the solar normal their direction of disappearance is and the longer filaments are, the more CMEs tend to be associated with them. In the future, we aim to complement the catalogue by using data from the CHAIN project (UeNo et al. 2014; Seki et al. 2018), as it archives the solar full-disk observation in the H-alpha line centre, red wing and blue wing during the night in Japan (while SDDI is unable to conduct observation).

**Acknowledgements**

We appreciate the daily routine observation, the maintenance and the development of the SDDI instruments by staffs of Hida Observatory. We also thank Dr. S. Yashiro for his great advices related to detection of filament-CME association. And we thank the referee for the useful comments, which greatly improved our manuscript. This work was supported by JSPS KAKENHI grant numbers JP15H05814 (Project for Solar-Terrestrial Environment

Prediction, PSTEP), JP16H03955, and JP18J23112. A. A. is financially supported by JSPJ KAKENHI grant number JP15H05816. D. S. is supported by Research Fellowships of Japan Society for the Promotion of Science for Young Scientists.

The Dst index used in this paper was provided by the WDC for Geomagnetism, Kyoto (http://wdc.kugi.kyoto-u.ac.jp/wdc/Sec3.html).

The SOHO/LASCO CME Catalog is generated and maintained at the CDAW Data Center by NASA and The Catholic University of America in cooperation with the Naval Research Laboratory. SOHO is a project of international cooperation between ESA and NASA.